\documentclass[aps, pra, notitlepage, onecolumn, floatfix, longbibliography]{revtex4-1}
 
\pdfoutput=1
\usepackage{graphicx}
\usepackage{amsmath, amssymb, amsfonts, bm}
\usepackage{bbm}
\usepackage{latexsym}
\usepackage[colorlinks,citecolor=blue]{hyperref}
\usepackage{MnSymbol}
\usepackage{verbatim}

\newcommand{\avg}[1]{\langle #1 \rangle}

\newcommand{\inpt}{\mathrm{in}}
\newcommand{\Floquet}{\mathrm{F}}
\newcommand{\Hc}{\mathrm{H.c.}}
\newcommand{\Hint}{\hat{H}_\mathrm{int}}
\newcommand{\rme}{{\rm e}}
\newcommand{\rmi}{{\rm i}}
\newcommand{\RWA}{RWA}

\begin{document}

\title{Combining Floquet and Lyapunov techniques for time-dependent problems in optomechanics and electromechanics}

\author{Iivari Pietik\"ainen}
\author{Ond\v{r}ej \v{C}ernot\'ik}
\author{Radim Filip}

\affiliation{Department of Optics, Palack\'y University, 17. listopadu 1192/12, 77146 Olomouc, Czechia}

\date{\today}

\begin{abstract}
	Cavity optomechanics and electromechanics form an established field of research investigating the interactions between electromagnetic fields and the motion of quantum mechanical resonators. In many applications, linearised form of the interaction is used, which allows for the system dynamics to be fully described using a Lyapunov equation for the covariance matrix of the Wigner function. This approach, however, is problematic in situations where the Hamiltonian becomes time dependent as is the case for systems driven at multiple frequencies simultaneously. This scenario is highly relevant as it leads to dissipative preparation of mechanical states or backaction-evading measurements of mechanical motion. The time-dependent dynamics can be solved with Floquet techniques whose application is, nevertheless, not straightforward. Here, we describe a general method for combining the Lyapunov approach with Floquet techniques that enables us to transform the initial time-dependent problem into a time-independent one, albeit in a larger Hilbert space. We show how the lengthy process of applying the Floquet formalism to the original equations of motion and deriving a Lyapunov equation from their time-independent form can be simplified with the use of properly defined Fourier components of the drift matrix of the original time-dependent system. We then use our formalism to comprehensively analyse dissipative generation of mechanical squeezing beyond the rotating wave approximation. Our method is applicable to various problems with multitone driving schemes in cavity optomechanics, electromechanics, and related disciplines.
\end{abstract}

\maketitle

\tableofcontents

\section{Introduction}

In the past years, cavity optomechanics and electromechanics~\cite{Aspelmeyer2014} became a mature field.
Starting with first experiments demonstrating ground-state cooling of mechanical motion~\cite{Chan2011,Teufel2011}, electromagnetic fields are now commonly used to manipulate and measure quantum states of mechanical resonators.
Examples of nonclassical mechanical states include squeezed states, prepared using reservoir engineering~\cite{Kronwald2013} or parametric effects~\cite{Liao2011}, as well as Fock states and their superpositions~\cite{Riedinger2016,Hong2017,Kounalakis2019}.
Preparation of these states presents a crucial first step towards creating more complex mechanical states such as truly macroscopic mechanical superpositions~\cite{Milburn2016,Shomroni2019,Zhan2019} or towards generation of entanglement in optomechanical and electromechanical systems~\cite{Tan2013,Woolley2014,Riedinger2018}.

In many applications, the system of interest is described by a Hamiltonian that is quadratic in the canonical operators of radiation and mechanical motion.
As the unitary dynamics are accompanied by single-excitation loss and gain and---in case of backaction-evading measurements---homodyne detection (nonunitary processes which are linear in the canonical operators), the phase-space description of the quantum state retains a Gaussian character~\cite{Weedbrook2012}.
We can therefore characterise the state by the first and second statistical moments of the quasiprobability density---the mean vector and covariance matrix of the canonical operators---instead of the density operator.
This presents a great advantage since we can formulate equations of motion for the mean and covariance matrix~\cite{Edwards2005,Cernotik2015} which can be solved more efficiently than the master equation that describes the evolution of the density operator.
Particularly the covariance matrix is important for investigating Gaussian quantum features of the resulting quantum state such as quadrature squeezing or entanglement~\cite{Weedbrook2012};
in the steady state, its form can be obtained by solving an algebraic equation of the Lyapunov (for unconditional dynamics) or Riccati (for conditional evolution) type~\cite{Edwards2005,Cernotik2015}.

A crucial obstacle for a more widespread application of these techniques is the explicit time dependence of the driving electromagnetic fields.
Dissipative preparation of mechanical states~\cite{Kronwald2013,Tan2013,Woolley2014,Li2015,Pirkkalainen2015,Wollman2015,Milburn2016,Pontin2016,Brunelli2018,Liao2018,Ockeloen-Korppi2018,Barzanjeh2019,Shomroni2019,Zhan2019} and tomographic backaction-evading measurements of mechanical motion~\cite{Ockeloen-Korppi2016,Delaney2019,Hertzberg2009,Shomroni2019b,Suh2014,Clerk2008,Lecocq2015,Lei2016,Brunelli2019} rely on driving the system with multiple fields at different frequencies while parametric squeezing requires modulation of the optical spring~\cite{Liao2011,Bothner2019,Cernotik2020,Chowdhury2020};
both of these approaches result in time-dependent optomechanical Hamiltonians.
The steady-state Lyapunov equation can then only be applied under the rotating wave approximation (\RWA{}) which neglects fast oscillating terms in the interaction and only keeps those that are resonant.
This approximation can work well as long as both the cavity damping rate and optomechanical coupling strength are much smaller than the mechanical frequency.
When one of these quantities becomes comparable to the mechanical frequency, the \RWA{} breaks down and the fast oscillating terms can no longer be neglected.
In this case, Floquet techniques can be used to find the steady state by expanding the system operators into their Fourier components~\cite{Mari2009,Malz2016,Malz2016a,Aranas2017,Qiu2019}.
Their application is, however, often cumbersome and lacks a simple, general approach.

In this article, we show how the Lyapunov and Floquet techniques can be combined to obtain a straightforward approach to finding the steady state of optomechanical and electromechanical systems with periodic Hamiltonians.
We derive a general method which can be directly applied to the Lyapunov equation of a time-dependent quantum system to obtain a time-independent Lyapunov equation in the larger Floquet space.
Importantly, the frequency components of the relevant quantities can be directly read off from the time-dependent equations of motion for the canonical operators;
this feature makes our strategy particularly easy to implement for a wide variety of critical problems in optomechanics and electromechanics.

After presenting the general method, we apply it to the problem of dissipative mechanical squeezing.
We consider both the usual squeezing obtained by cooling a mechanical Bogoliubov mode~\cite{Kronwald2013} as well as the recently proposed combination of dissipative and parametric squeezing possible for levitated particles~\cite{Cernotik2020} and analyse the resulting squeezing in both cases beyond the usual \RWA{}.
This approximation is invalidated not only by the nonzero sideband ratio (the ratio between the cavity decay rate and the mechanical frequency) but also by the ultrastrong coupling regime which is being reached in an evergrowing number of optomechanical and electromechanical systems~\cite{Heikkila2014,Pirkkalainen2015a,Romero-Sanchez2018,Delic2019,Fogliano2019,Peterson2019,Rodrigues2019,Schmidt2019,Windey2019,Delic2020}.
A detailed systematic analysis of the effects that the counterrotating terms will thus have important consequences not only for immediate applications in state-of-the-art optomechanical and electromechanical systems but also for advanced state preparation techniques, possibly going beyond the Gaussian regime.

\section{Gaussian optomechanics and electromechanics}

\subsection{Lyapunov equation}

The dynamics of an open quantum system can be described by the master equation ($\hbar = 1$)
\begin{equation}\label{eq:MEq}
\dot{\hat{\rho}} = -\rmi[\hat{H},\hat{\rho}] +\sum_n \mathcal{D}[\hat{\jmath}_n]\hat{\rho} \,,
\end{equation}
where $\mathcal{D}[\hat{\jmath}]\hat\rho = \hat{\jmath}\hat\rho \hat{\jmath}^\dagger -\frac{1}{2}\hat{\jmath}^\dagger\hat{\jmath}\hat\rho -\frac{1}{2}\hat\rho\hat{\jmath}^\dagger\hat{\jmath}$ is the Lindblad superoperator with collapse operator $\hat{\jmath}$~\cite{Wiseman2010}.
The full dynamics can, in principle, be solved from this equation but this is not always feasible in practice.
Optomechanical systems are one such example; the large thermal occupation of the mechanical bath requires prohibitively large Hilbert-space dimensions in the Fock-state basis to be considered for numerical simulations.

Full description of the dynamics without solving the master equation~\eqref{eq:MEq} is possible for Gaussian systems,
which are characterised by Gaussian quasiprobability distributions in phase space and as such can be fully described by the first and second statistical moments of these distributions.
A system is Gaussian when its Hamiltonian $\hat{H}$ is quadratic and the collapse operators $\hat{\jmath}_n$ linear in canonical operators; then, we can solve for the covariance matrix describing the state of the system instead of the density matrix.
(Note that the first moments remain zero when the Hamiltonian contains no linear terms and the collapse operators no constant terms.)

The covariance matrix $\mathbf{\Gamma}$ obeys the Lyapunov equation
\begin{equation}\label{eq:Lyapunov}
\dot{\mathbf{\Gamma}} = \mathbf{A} \mathbf{\Gamma} +\mathbf{\Gamma A}^\top + \mathbf{N},
\end{equation}
where
\begin{subequations}
\begin{align}
\Gamma_{ij} &= \langle \hat{r}_i \hat{r}_j +\hat{r}_j \hat{r}_i \rangle -2\langle \hat{r}_i\rangle\langle \hat{r}_j\rangle ,\\
\mathbf{\hat{r}} &= \left( \hat{q}_1, \hat{p}_1, \hat{q}_2, \hat{p}_2,\ldots, \hat{q}_N, \hat{p}_N \right)^\top,
\end{align}
\end{subequations}
with the canonical commutation relation
\begin{equation}\label{eq:CR}
	[\mathbf{\hat{r}},\mathbf{\hat{r}}^\top] = \rmi\sigma_N = \rmi(\underbrace{\sigma \oplus \sigma \oplus \ldots \oplus \sigma}_N),\qquad 
	\sigma = \left(\begin{array}{cc} 0&1 \\ -1&0 \end{array}\right),
\end{equation}
where $\sigma$ is the one-mode symplectic matrix.
The drift ($\mathbf{A}$) and diffusion ($\mathbf{N}$) matrices can be found from the master equation~\cite{Cernotik2015} or from the Langevin equations
\begin{equation}\label{eq:dr}
\mathbf{\dot{\hat{r}}} = \mathbf{A}\mathbf{\hat{r}} +\hat{\xi},
\end{equation}
where $\hat{\xi} = (\hat{q}_{1,\inpt},\hat{p}_{1,\inpt},\ldots,\hat{q}_{N,\inpt},\hat{p}_{N,\inpt})^\top$ is the vector of noise operators.
The diffusion matrix can be found from the correlation properties of the noise via
\begin{equation}
	\mathbf{N} = \avg{\hat{\xi}(t)\hat{\xi}(t)^\top}.
\end{equation}

The Lyapunov equation~\eqref{eq:Lyapunov} is a powerful tool in optomechanics where it can be used to evaluate the dynamics of the system or to find its steady state from the algebraic form of the Lyapunov equation
\begin{equation}\label{eq:LyapunovSS}
\mathbf{A} \mathbf{\Gamma} +\mathbf{\Gamma A}^\top + \mathbf{N} = 0.
\end{equation}
The process of solving Eq.~\eqref{eq:Lyapunov} or~\eqref{eq:LyapunovSS} is straightforward as long as the system operators are not explicitly time dependent.
When the Hamiltonian or coupling to the environment becomes time dependent, the drift matrix (and possibly also the diffusion matrix) becomes time dependent as well.
While the differential Lyapunov equation~\eqref{eq:Lyapunov} can still be solved in this case, a straightforward solution of the algebraic Lyapunov equation~\eqref{eq:LyapunovSS} is not possible.

\subsection{The Floquet--Lyapunov method}\label{ssec:Floquet}

A common approach to problems with periodic time-dependence is based on the Floquet formalism,
which allows us to transform a set of periodic linear differential equations into a larger set of linear differential equations without explicit time dependence~\cite{Shirley1965}. Here, we will show how to use this method to obtain a time-independent Lyapunov equation for explicitly time-dependent dynamics.
For simplicity, we will assume that only the Hamiltonian is $\tau$-periodic, $\hat{H}(t+\tau) = \hat{H}(t)$, leading to a periodic drift matrix $\mathbf{A}(t+\tau) = \mathbf{A}(t)$, while the coupling to the environment (and the diffusion matrix $\mathbf{N}$) remains explicitly time independent.
We will only present a general recipe for obtaining a time-independent Lyapunov equation in this section;
the mathematical justification of these steps is presented in Appendix~\ref{app:Floquet}.

We start by expressing the operators and the drift matrix in terms of their Fourier components at frequency $\omega = 2\pi/\tau$,
\begin{subequations}\label{eq:Fourier}
\begin{align}
	\mathbf{\hat{r}} &= \mathbf{\hat{r}}^{(0)} +\sqrt{2}\sum_{n=1}^\infty [\mathbf{\hat{r}}_\mathrm{c}^{(n)}\cos(n\omega t) + \mathbf{\hat{r}}_\mathrm{s}^{(n)}\sin(n\omega t)] \,,\\
	\mathbf{A}(t) &= \mathbf{A}^{(0)} +\sqrt{2}\sum_{n=1}^\infty [\mathbf{A}_\mathrm{c}^{(n)}\cos(n\omega t) + \mathbf{A}_\mathrm{s}^{(n)}\sin(n\omega t)] \,,\\
	{\hat{\xi}} &= {\hat{\xi}}^{(0)} +\sqrt{2}\sum_{n=1}^\infty [{\hat{\xi}}_\mathrm{c}^{(n)}\cos(n\omega t) + {\hat{\xi}}_\mathrm{s}^{(n)}\sin(n\omega t)] \,,
\end{align}
\end{subequations}
The frequency components $\mathbf{\hat{r}}^{(0)}, \mathbf{\hat{r}}_\mathrm{c,s}^{(n)}$ fulfil the canonical commutation relation, $[\hat{\mathbf{r}}^{(0)},\hat{\mathbf{r}}^{(0)\top}] = [\hat{\mathbf{r}}_{\rm k}^{(n)},\hat{\mathbf{r}}_{\rm k}^{(n)\top}] = \rmi\sigma_N$ (here, ${\rm k}\in\{\rm{c,s}\}$ and operators from different frequency components commute), and obey the Langevin equations
\begin{subequations}\label{eq:FTLangevin}
\begin{align}
	\mathbf{\dot{\hat{r}}}^{(0)} &= \mathbf{A}^{(0)}\mathbf{\hat{r}}^{(0)} + \sum_{m=1}^\infty(\mathbf{A}_\mathrm{c}^{(m)}\mathbf{\hat{r}}_\mathrm{c}^{(m)} + \mathbf{A}_\mathrm{s}^{(m)}\mathbf{\hat{r}}_\mathrm{s}^{(m)}) + \hat{\xi}^{(0)}, \\
	\begin{split}
	\mathbf{\dot{\hat{r}}}_\mathrm{c}^{(n)} &= \mathbf{A}^{(0)}\mathbf{\hat{r}}_\mathrm{c}^{(n)} +\mathbf{A}_\mathrm{c}^{(n)}\mathbf{\hat{r}}^{(0)} -n\omega\mathbf{r}_\mathrm{s}^{(n)}
		+\frac{1}{\sqrt{2}}\sum_{m=1}^\infty \mathbf{A}_\mathrm{c}^{(m)}(\mathbf{\hat{r}}_\mathrm{c}^{(m-n)}+\mathbf{\hat{r}}_\mathrm{c}^{(m+n)} +\mathbf{\hat{r}}_\mathrm{c}^{(n-m)}) \\
		&\quad +\frac{1}{\sqrt{2}}\sum_{m=1}^\infty \mathbf{A}_\mathrm{s}^{(m)}(\mathbf{\hat{r}}_\mathrm{s}^{(m-n)}+\mathbf{\hat{r}}_\mathrm{s}^{(m+n)} -\mathbf{\hat{r}}_\mathrm{s}^{(n-m)}) +\hat{\xi}_\mathrm{c}^{(n)},
		\end{split}\\
	\begin{split}
	\mathbf{\dot{\hat{r}}}_\mathrm{s}^{(n)} &= \mathbf{A}^{(0)}\mathbf{\hat{r}}_\mathrm{s}^{(n)} +\mathbf{A}_\mathrm{s}^{(n)}\mathbf{\hat{r}}^{(0)} +n\omega\mathbf{r}_\mathrm{c}^{(n)}
		+\frac{1}{\sqrt{2}}\sum_{m=1}^\infty \mathbf{A}_\mathrm{c}^{(m)}(-\mathbf{\hat{r}}_\mathrm{s}^{(m-n)}+\mathbf{\hat{r}}_\mathrm{s}^{(m+n)} +\mathbf{\hat{r}}_\mathrm{s}^{(n-m)}) \\
		&\quad +\frac{1}{\sqrt{2}}\sum_{m=1}^\infty \mathbf{A}_\mathrm{s}^{(m)}(\mathbf{\hat{r}}_\mathrm{c}^{(m-n)}-\mathbf{\hat{r}}_\mathrm{c}^{(m+n)} +\mathbf{\hat{r}}_\mathrm{c}^{(n-m)})+\hat{\xi}_\mathrm{s}^{(n)}.
		\end{split}
\end{align}
\end{subequations}
The individual frequency components of the noise operators have the same correlation properties as the initial noise, $\avg{\hat{\xi}^{(0)}\hat{\xi}^{(0)\top}} = \avg{\hat{\xi}_{\rm k}^{(n)}\hat{\xi}_{\rm k}^{(n)\top}} = \mathbf{N}$,
and are mutually uncorrelated, $\avg{\hat{\xi}^{(0)}\hat{\xi}_{\rm k}^{(n)\top}} =\avg{\hat{\xi}_{\rm k}^{(n)}\hat{\xi}_{\rm l}^{(m)\top}} = 0$ for $n\neq m$ or ${\rm k}\neq\rm{ l}$.
Note also that in Eqs.~\eqref{eq:FTLangevin}, only the frequency components with $\pm m\pm n > 0$ should be included when evaluating the sums; effectively, we have $\mathbf{\hat{r}}_\mathrm{k}^{(n)} = 0$ for $n\leq 0$.

Collecting the frequency components in the vector $\mathbf{\hat{r}}_\Floquet = (\mathbf{\hat{r}}^{(0)},\mathbf{\hat{r}}_\mathrm{c}^{(1)},\mathbf{\hat{r}}_\mathrm{s}^{(1)},\mathbf{\hat{r}}_\mathrm{c}^{(2)},\mathbf{\hat{r}}_\mathrm{s}^{(2)},\ldots)^\top$ (the components $\mathbf{\hat{r}}^{(0)},\mathbf{\hat{r}}_{\rm k}^{(n)}$, which are column vectors, are arranged into a longer column vector),
we can formulate the Langevin equation
\begin{equation}
\mathbf{\dot{\hat{r}}}_\Floquet = \mathbf{A}_\Floquet \mathbf{\hat{r}}_\Floquet + \hat{\xi}_\Floquet \,,
\end{equation}
where we introduced the drift matrix
\begin{equation}\label{eq:Ar}
\mathbf{A}_\Floquet = 
\begin{pmatrix}
\mathbf{A}^{(0)} &\mathbf{A}_{\rm c}^{(1)} &\mathbf{A}_{\rm s}^{(1)} &\mathbf{A}_{\rm c}^{(2)} &\mathbf{A}_{\rm s}^{(2)} & \\
\mathbf{A}_{\rm c}^{(1)} &\mathbf{A}^{(0)} +\frac{1}{\sqrt{2}}\mathbf{A}_{\rm c}^{(2)} &-\omega\mathbf{I}_N +\frac{1}{\sqrt{2}}\mathbf{A}_{\rm s}^{(2)} &\frac{1}{\sqrt{2}}(\mathbf{A}_{\rm c}^{(1)} +\mathbf{A}_{\rm c}^{(3)}) &\frac{1}{\sqrt{2}}(\mathbf{A}_{\rm s}^{(1)} +\mathbf{A}_{\rm s}^{(3)}) & \\
\mathbf{A}_{\rm s}^{(1)} &\omega\mathbf{I}_N +\frac{1}{\sqrt{2}}\mathbf{A}_{\rm s}^{(2)} &\mathbf{A}^{(0)} -\frac{1}{\sqrt{2}}\mathbf{A}_{\rm c}^{(2)} &\frac{1}{\sqrt{2}}(-\mathbf{A}_{\rm s}^{(1)} +\mathbf{A}_{\rm s}^{(3)}) &\frac{1}{\sqrt{2}}(\mathbf{A}_{\rm c}^{(1)} -\mathbf{A}_{\rm c}^{(3)}) &\ldots \\
\mathbf{A}_{\rm c}^{(2)} &\frac{1}{\sqrt{2}}(\mathbf{A}_{\rm c}^{(1)} +\mathbf{A}_{\rm c}^{(3)}) &\frac{1}{\sqrt{2}}(-\mathbf{A}_{\rm s}^{(1)} +\mathbf{A}_{\rm s}^{(3)}) &\mathbf{A}^{(0)} +\frac{1}{\sqrt{2}}\mathbf{A}_{\rm c}^{(4)} &-2\omega\mathbf{I}_N +\frac{1}{\sqrt{2}}\mathbf{A}_{\rm s}^{(4)} & \\
\mathbf{A}_{\rm s}^{(2)} &\frac{1}{\sqrt{2}}(\mathbf{A}_{\rm s}^{(1)} +\mathbf{A}_{\rm s}^{(3)}) &\frac{1}{\sqrt{2}}(\mathbf{A}_{\rm c}^{(1)} -\mathbf{A}_{\rm c}^{(3)}) &2\omega\mathbf{I}_N +\frac{1}{\sqrt{2}}\mathbf{A}_{\rm s}^{(4)}  &\mathbf{A}^{(0)} -\frac{1}{\sqrt{2}}\mathbf{A}_{\rm c}^{(4)} & \\
 & &\vdots & & &\ddots
\end{pmatrix}\,,
\end{equation}with $\mathbf{I}_N$ denoting the $N$-mode (i.e., $2N$-dimensional) identity matrix
and $\hat{\xi}_\Floquet = (\hat{\xi}^{(0)},\hat{\xi}_\mathrm{c}^{(1)},\hat{\xi}_\mathrm{s}^{(1)},\ldots)^\top$ being the vector of noise operators in terms of their frequency components.
In these expressions, we use the subscript F to signify that the quantities are defined in the Floquet space.
We can now formulate the Lyapunov equation in the Floquet space
\begin{equation}\label{eq:FloquetLyap}
	\dot{\mathbf{\Gamma}}_\Floquet = \mathbf{A}_\Floquet\mathbf{\Gamma}_\Floquet + \mathbf{\Gamma}_\Floquet\mathbf{A}_\Floquet^\top + \mathbf{N}_\Floquet
\end{equation}
with the time-independent drift matrix given by Eq.~\eqref{eq:Ar} and diffusion matrix $\mathbf{N}_\Floquet = {\rm diag}(\mathbf{N},\mathbf{N},...)$.
The covariance matrix can be expressed in the block form
\begin{equation}
	\mathbf{\Gamma}_\Floquet = 
	\begin{pmatrix}
		\mathbf{\Gamma}^{(0)} & \mathbf{\Gamma}_{\rm c}^{(01)} & \mathbf{\Gamma}_{\rm s}^{(01)} \\
		\mathbf{\Gamma}_{\rm c}^{(10)} & \mathbf{\Gamma}_{\rm c}^{(1)} & \mathbf{\Gamma}_{\rm cs}^{(11)} & \ldots \\
		\mathbf{\Gamma}_{\rm s}^{(10)} & \mathbf{\Gamma}_{\rm sc}^{(11)} & \mathbf{\Gamma}_{\rm s}^{(1)} \\
		& \vdots & &\ddots
	\end{pmatrix},
\end{equation}
where the blocks are defined as the covariances between the corresponding frequency components of the quadrature operators;
we thus have
\begin{align}
\begin{split}
	\Gamma_{ij}^{(0)} = \avg{\hat{r}_i^{(0)}\hat{r}_j^{(0)} + \hat{r}_j^{(0)}\hat{r}_i^{(0)}} - 2\avg{\hat{r}_i^{(0)}}\avg{\hat{r}_j^{(0)}},\qquad &
	\Gamma_{{\rm k},ij}^{(0n)} = \avg{\hat{r}_i^{(0)}\hat{r}_{{\rm k},j}^{(n)} + \hat{r}_{{\rm k},j}^{(n)}\hat{r}_i^{(0)}} - 2\avg{\hat{r}_i^{(0)}}\avg{\hat{r}_{{\rm k},j}^{(n)}}, \\
	\Gamma_{{\rm k},ij}^{(n)} = \avg{\hat{r}_{{\rm k},i}^{(n)}\hat{r}_{{\rm k},j}^{(n)} + \hat{r}_{{\rm k},j}^{(n)}\hat{r}_{{\rm k},i}^{(n)}} - 2\avg{\hat{r}_{{\rm k},i}^{(n)}}\avg{\hat{r}_{{\rm k},j}^{(n)}},\qquad &
	\Gamma_{{\rm kl},ij}^{(mn)} = \avg{\hat{r}_{{\rm k},i}^{(m)}\hat{r}_{{\rm l},j}^{(n)} + \hat{r}_{{\rm l},j}^{(n)}\hat{r}_{{\rm k},i}^{(m)}} - 2\avg{\hat{r}_{{\rm k},i}^{(m)}}\avg{\hat{r}_{{\rm l},j}^{(n)}}.
\end{split}
\end{align}
The steady state can now be calculated from Eq.~\eqref{eq:FloquetLyap} by setting $\dot{\mathbf{\Gamma}}_\Floquet = 0$, which gives a linear, time-independent algebraic equation.

The system now becomes effectively infinite dimensional and must be suitably truncated for numerical simulations.
This truncation has to ensure that the solution (contained in the zeroth Brillouin zone, \emph{i.e.}, the covariance block $\mathbf{\Gamma}^{(0)}$) converges.
As we shall show in the following examples, only a small number of Brillouin zones is typically sufficient, making our approach more feasible than direct solution of the master equation or finding the long-time limit of the differential Lyapunov equation with time-dependent drift.

\subsection{Illustration: optomechanical cooling}

To illustrate our formalism, we now consider optomechanical sideband cooling and show how the Floquet--Lyapunov techniques can be used to obtain the correct steady-state mechanical occupation in the rotating frame.
For an optical cavity (with annihilation operator $\hat{c}$) coupled to a mechanical oscillator (annihilation operator $\hat{b}$), the linearised optomechanical Hamiltonian (in the lab frame of the mechanical motion) takes the form~\cite{Aspelmeyer2014}
\begin{equation}\label{eq:HamOC}
	\hat{H} = \Delta\hat{c}^\dagger\hat{c} +\omega_\mathrm{m}\hat{b}^\dagger\hat{b} + g(\hat{c}^\dagger +\hat{c})(\hat{b}^\dagger +\hat{b}) \,,
\end{equation}
where $\Delta = \omega_\mathrm{c}-\omega_\mathrm{L}$ is the detuning between the cavity frequency $\omega_\mathrm{c}$ and driving frequency $\omega_\mathrm{L}$, $\omega_\mathrm{m}$ is the mechanical frequency, and $g$ is the coupling constant.
This is a time-independent Hamiltonian and the dynamics can be solved directly by using the Lyapunov equation.

\begin{figure}
  \centering
  \includegraphics[width=0.7\linewidth]{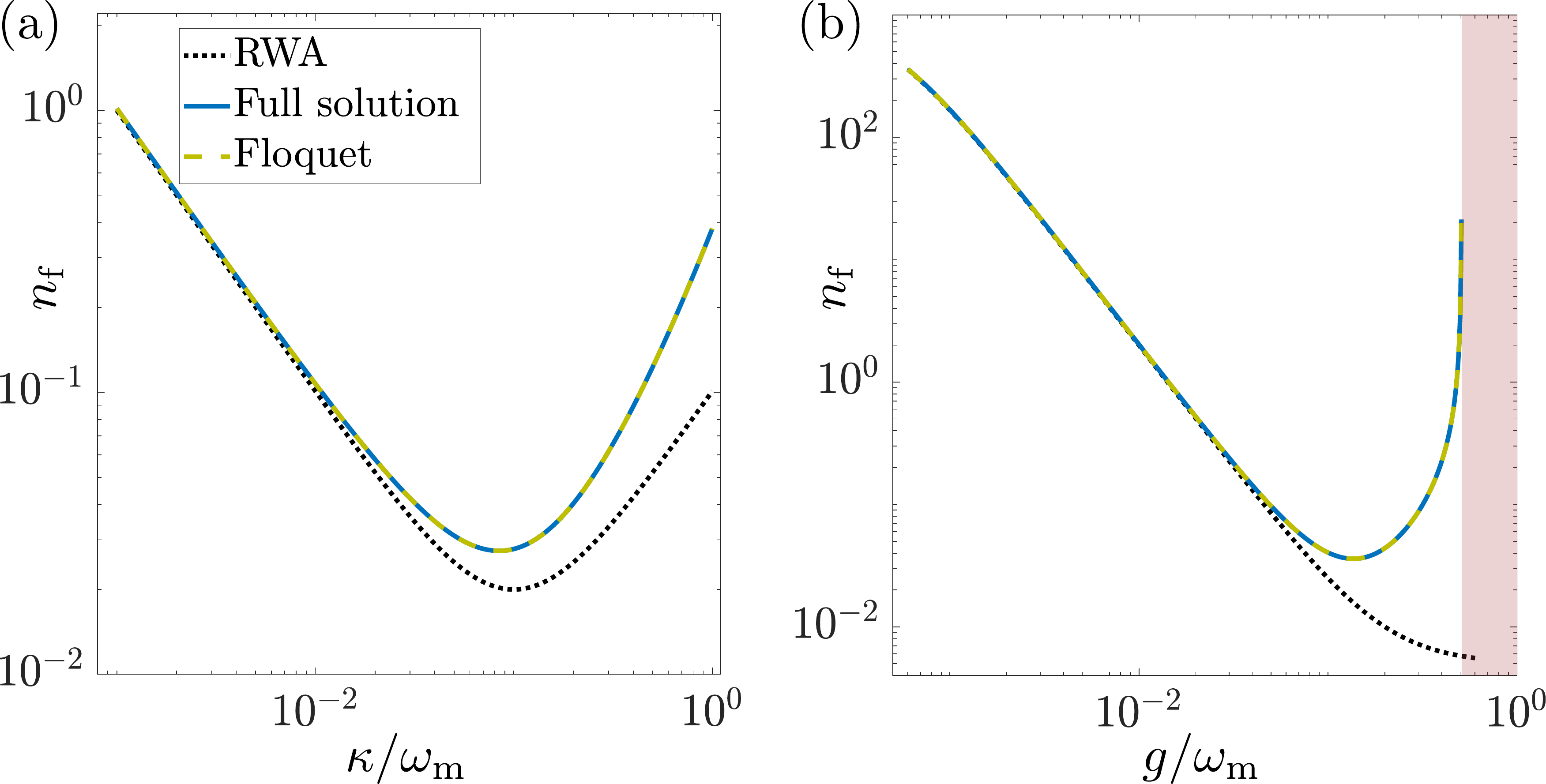}
  \caption{Mechanical population as a function of (a) cavity dissipation and (b) coupling strength. The steady-state mechanical population is calculated by solving the Lyapunov equation using the full drift matrix (Eq.~(\ref{eq:LyapCooling}), blue solid line), under the \RWA{} (Eq.~(\ref{eq:RWAA}), black dotted line), and using the Floquet--Lyapunov approach (Eq.~(\ref{eq:FloqCooling}), yellow dashed line). In the red region the system is unstable. The parameters used for the simulations are $g/\omega_{\rm m}=0.1$, $\kappa/\omega_{\rm m}=0.2$, $\gamma/\omega_{\rm m}=10^{-6}$, and $\bar{n}=10^3$.}\label{fig:OC1}
\end{figure}

Introducing the canonical operators $\hat{q}_1 = (\hat{c}+\hat{c}^\dagger)/\sqrt{2}$, $\hat{p}_1 = -\rmi(\hat{c}-\hat{c}^\dagger)/\sqrt{2}$ for the cavity field and $\hat{q}_2$, $\hat{p}_2$ for the mechanics (defined similarly), we can describe the dynamics using the Langevin equations
\begin{subequations}
\begin{align}
	\dot{\hat{q}}_1 &= -\kappa\hat{q}_1 + \Delta\hat{p}_1 + \sqrt{2\kappa}\hat{q}_{1,\inpt},\\
	\dot{\hat{p}}_1 &= -\Delta\hat{q}_1 - \kappa\hat{p}_1 - 2g\hat{q}_2 + \sqrt{2\kappa}\hat{p}_{1,\inpt},\\
	\dot{\hat{q}}_2 &= -\gamma\hat{q}_2 + \omega_\mathrm{m}\hat{p}_2 + \sqrt{2\gamma}\hat{q}_{2,\inpt},\\
	\dot{\hat{p}}_2 &= - 2g\hat{q}_1 - \omega_\mathrm{m}\hat{q}_2 - \gamma\hat{p}_2 + \sqrt{2\gamma}\hat{p}_{2,\inpt}.
\end{align}
\end{subequations}
Here, $\kappa$ and $\gamma$ are the damping rates for the cavity field and the mechanical mode and the input noise operators of the optical and mechanical baths have the correlations 
\begin{subequations}
\begin{align}
	\avg{\hat{q}_{1,\inpt}(t)\hat{q}_{1,\inpt}(t')} = \avg{\hat{p}_{1,\inpt}(t)\hat{p}_{1,\inpt}(t')} &= \delta(t-t'),\\
	\avg{\hat{q}_{2,\inpt}(t)\hat{q}_{2,\inpt}(t')} = \avg{\hat{p}_{2,\inpt}(t)\hat{p}_{2,\inpt}(t')} &= (2\bar{n}+1)\delta(t-t')
\end{align}	
\end{subequations}
with $\bar{n}$ denoting the average occupation of the thermal mechanical bath;
all other correlations are zero.
From these equations, we obtain the Lyapunov equation
\begin{subequations}\label{eq:LyapCooling}
\begin{align}
	\dot{\mathbf{\Gamma}} &= \mathbf{A\Gamma} + \mathbf{\Gamma A}^\top + \mathbf{N},\\
	\mathbf{A} &= 
\begin{pmatrix}
-\kappa &\Delta &0 &0 \\
-\Delta &-\kappa & -2g &0 \\
0 &0 &-\gamma &\omega_\mathrm{m} \\
-2g &0 &-\omega_\mathrm{m} &-\gamma
\end{pmatrix} \,,\\
\mathbf{N} &= {\rm diag}[2\kappa,2\kappa,2\gamma(2\bar{n}+1),2\gamma(2\bar{n}+1)] \,.
\end{align}
\end{subequations}
The optimal cooling performance is then achieved for driving on the lower mechanical sideband, $\Delta = \omega_\mathrm{m}$, and with a sideband-resolved system, $\kappa\ll\omega_\mathrm{m}$~\cite{Marquardt2007,Wilson-Rae2007}.
This is due to the passive beam-splitter part of the optomechanical interaction being resonantly enhanced while the active two-mode-squeezing part (responsible for depositing energy into the mechanical mode) is far off resonant.
The steady-state mechanical occupation, obtained by setting $\dot{\mathbf{\Gamma}} = 0$ in Eq.~\eqref{eq:LyapCooling}, can be found as $n_\mathrm{f} = \frac{1}{4}(\Gamma_{33}+\Gamma_{44}-2)$ and is plotted Fig.~\ref{fig:OC1} as the solid blue line.

To demonstrate the Floquet formalism, we start from the Hamiltonian (\ref{eq:HamOC}) driven on the red sideband, $\Delta = \omega_{\rm m}$, and move to the interaction picture with respect to $\hat{H}_0 = \omega_\mathrm{m}(\hat{c}^\dagger \hat{c} +\hat{b}^\dagger\hat{b})$. Thus, the Hamiltonian becomes
\begin{equation}\label{eq:HamIntPic}
\hat{H} = g(\hat{c}^\dagger\hat{b} +\hat{c}\hat{b}^\dagger) +g(\hat{c}^\dagger\hat{b}^\dagger \rme^{2\rmi\omega_\mathrm{m} t} +\hat{c}\hat{b}\,\rme^{-2\rmi\omega_\mathrm{m} t}) \,.
\end{equation}
By applying the \RWA{} (valid for high mechanical frequencies, $\omega_{\rm m}\gg\kappa,g$), we are left with the beam-splitter Hamiltonian
\begin{equation}\label{eq:HamRWA}
\hat{H} = g(\hat{c}^\dagger\hat{b} +\hat{c}\hat{b}^\dagger) \,.
\end{equation}
This results in the drift matrix
\begin{equation}\label{eq:RWAA}
\mathbf{A} =
\begin{pmatrix}
-\kappa &0 &0 & g \\
0 &-\kappa & -g &0 \\
0 & g &-\gamma &0 \\
-g &0 &0 &-\gamma
\end{pmatrix} \,,
\end{equation}
while the diffusion matrix $\mathbf{N}$ remains the same as before.
The \RWA{} clearly demonstrates the mechanism of cooling---mechanical excitations are swapped to the cavity field from which they leak out through the cavity mirrors---but does not properly estimate the final mechanical occupation since it does not include the effective temperature of the optical bath induced by the two-mode squeezing part of the optomechanical interaction.
This is visible in regimes where the cavity damping or the optomechanical coupling start approaching the mechanical frequency, invalidating the assumptions necessary for the \RWA{} (see the black dotted line in Fig.~\ref{fig:OC1}, calculated in the same way as for the full model).

The full rotating-frame Hamiltonian \eqref{eq:HamIntPic} is, however, periodic with frequency $2\omega_\mathrm{m}$ so we can include the effect of the counterrotating terms in our simulations by applying the Floquet--Lyapunov approach.
We begin by forming the equations of motion for the canonical operators,
\begin{subequations}
\begin{align}
	\dot{\hat{q}}_1 &= -\kappa\hat{q}_1 + g\hat{p}_2 - g\hat{p}_2\cos(2\omega_\mathrm{m}t) +g\hat{q}_2\sin(2\omega_\mathrm{m}t) +\sqrt{2\kappa}\,\hat{q}_{1,\inpt}, \\
	\dot{\hat{p}}_1 &= -\kappa\hat{p}_1 - g\hat{q}_2 - g\hat{q}_2\cos(2\omega_\mathrm{m}t) -g\hat{p}_2\sin(2\omega_\mathrm{m}t) +\sqrt{2\kappa}\,\hat{p}_{1,\inpt}, \\
	\dot{\hat{q}}_2 &= -\gamma\hat{q}_2 + g\hat{p}_1 - g\hat{p}_1\cos(2\omega_\mathrm{m}t) +g\hat{q}_1\sin(2\omega_\mathrm{m}t) +\sqrt{2\gamma}\,\hat{q}_{2,\inpt}, \\
	\dot{\hat{p}}_2 &= -\gamma\hat{p}_2 - g\hat{q}_1 - g\hat{q}_1\cos(2\omega_\mathrm{m}t) -g\hat{p}_1\sin(2\omega_\mathrm{m}t) +\sqrt{2\gamma}\,\hat{p}_{2,\inpt}.
\end{align}
\end{subequations}
From these equations, we can readily read off the frequency components of the Floquet-space drift matrix,
\begin{equation}
	\mathbf{A}^{(0)} = \mathbf{A}_0 + g\mathbf{J}_-,\qquad \mathbf{A}_{\rm c}^{(1)} = -\frac{g}{\sqrt{2}}\mathbf{J}_+,\qquad \mathbf{A}_{\rm s}^{(1)} = \frac{g}{\sqrt{2}}\mathbf{G}_1,
\end{equation}
where $\mathbf{A}_0 = {\rm diag}(-\kappa,-\kappa,-\gamma,-\gamma)$ is the part of the drift matrix associated with damping,
\begin{equation}
	\mathbf{J}_\pm = \begin{pmatrix} 0 & 0 & 0 & 1 \\ 0 & 0 & \pm 1 & 0 \\ 0 & 1 & 0 & 0 \\ \pm 1 & 0 & 0 & 0
	\end{pmatrix},
	\qquad {\rm and}\qquad 
	\mathbf{G}_1 = \begin{pmatrix}
		0 & 0 & 1 & 0 \\
		0 & 0 & 0 & -1 \\
		1 & 0 & 0 & 0 \\
		0 & -1 & 0 & 0
	\end{pmatrix}.
\end{equation}
Keeping the first three Brillouin zones (i.e., defining the Floquet vector $\mathbf{\hat{r}}_\Floquet = (\mathbf{\hat{r}}^{(0)},\mathbf{\hat{r}}_\mathrm{c}^{(1)},\mathbf{\hat{r}}_\mathrm{s}^{(1)})^\top$), we have the Lyapunov equation
\begin{subequations}\label{eq:FloqCooling}
\begin{align}
	\dot{\mathbf{\Gamma}}_\Floquet &= \mathbf{A}_\Floquet\mathbf{\Gamma}_\Floquet +\mathbf{\Gamma}_\Floquet\mathbf{A}_\Floquet^\top+\mathbf{N}_\Floquet,\\
	\mathbf{A}_\Floquet &= \begin{pmatrix}
		\mathbf{A}^{(0)} & \mathbf{A}_\mathrm{c}^{(1)} & \mathbf{A}_\mathrm{s}^{(1)} \\
		\mathbf{A}_\mathrm{c}^{(1)} & \mathbf{A}^{(0)} & -2\omega_\mathrm{m}\mathbf{I}_2 \\
		\mathbf{A}_\mathrm{s}^{(1)} & 2\omega_\mathrm{m}\mathbf{I}_2 & \mathbf{A}^{(0)}
	\end{pmatrix}, \\
	\mathbf{N}_\Floquet &= \mathrm{diag}(\mathbf{N},\mathbf{N},\mathbf{N}).
\end{align}
\end{subequations}
The steady-state mechanical occupation (dashed yellow line in Fig.~\ref{fig:OC1}) now shows perfect agreement with the full model.

\section{Dissipative squeezing beyond the rotating wave approximation}

In the previous example of optomechanical cooling, the use of the Floquet method is unnecessary as the system can be represented in a frame where the Hamiltonian is time independent.
In the rest of this paper, we focus on more complicated driving schemes for which the optomechanical system remains time-dependent regardless of the reference frame.
This is generally the case when the system is driven with multiple fields at different frequencies or, equivalently, using a single drive with a modulated amplitude.
We focus on the simplest nontrivial example---one mechanical mode and one cavity field subject to a two-tone drive.
We perform a detailed analysis of the mechanical squeezing that can be achieved in this system originally proposed by Kronwald \emph{et al.}~\cite{Kronwald2013} in section~\ref{ssec:DisSq}.
In section~\ref{ssec:levitation}, we then analyse a recently proposed scheme for generating steady-state mechanical squeezing in levitated systems by a combination of parametric and dissipative squeezing~\cite{Cernotik2020}.

\subsection{Mechanical squeezing with a  two-tone drive}\label{ssec:DisSq}

Following Ref.~\cite{Kronwald2013}, we start from the full optomechanical Hamiltonian under two-tone driving,
\begin{equation}
	\hat{H} = \omega_\mathrm{c}\hat{c}^\dagger\hat{c}+\omega_\mathrm{m}\hat{b}^\dagger\hat{b} +g_0\hat{c}^\dagger\hat{c}(\hat{b}+\hat{b}^\dagger) +(\eta_+\rme^{-\rmi\omega_+t}+\eta_-\rme^{-\rmi\omega_-t})\hat{c}^\dagger +\Hc,
\end{equation}
where $g_0$ is the single-photon coupling strength and $\eta_\pm$ are the amplitudes of the drives at frequencies $\omega_\pm = \omega_\mathrm{c}\pm\omega_\mathrm{m}$.
Under this driving, the cavity field acquires a large classical amplitude $\alpha_+\rme^{-\rmi\omega_+t}+\alpha_-\rme^{-\rmi\omega_-t}$;
introducing the linearised coupling rates $g_\pm = g_0\alpha_\pm$ (assuming, for simplicity, $\alpha_\pm\in\mathbb{R}$) and moving to the interaction picture with respect to $\hat{H}_0 = \omega_-\hat{c}^\dagger\hat{c}+\omega_\mathrm{m}\hat{b}^\dagger\hat{b}$, we obtain the interaction Hamiltonian
\begin{equation}\label{eq:HamDS}
\hat{H} = g_-(\hat{c}^\dagger\hat{b}+\hat{b}^\dagger\hat{c}) + g_+(\hat{c}\hat{b}+\hat{c}^\dagger\hat{b}^\dagger) 
	+g_-(\hat{c}\hat{b}\,\rme^{-2\rmi\omega_\mathrm{m}t}+\hat{c}^\dagger\hat{b}^\dagger \rme^{2\rmi\omega_\mathrm{m}t})
	+g_+(\hat{c}^\dagger\hat{b}\rme^{-2\rmi\omega_\mathrm{m}t} +\hat{b}^\dagger\hat{c}\,\rme^{2\rmi\omega_\mathrm{m}t}).
\end{equation}

To follow the Floquet approach, we first formulate the equations of motion
\begin{subequations}
\begin{eqnarray}
\dot{\hat{q}}_1 &=& -\kappa\hat{q}_1 +(g_- -g_+)\hat{p}_2 -(g_- -g_+)\hat{p}_2\cos(2\omega_{\rm m} t) +(g_- -g_+)\hat{q}_2\sin(2\omega_{\rm m} t)+\sqrt{2\kappa}\hat{q}_{1,\inpt}, \\
\dot{\hat{p}}_1 &=& -\kappa\hat{p}_1 -(g_- +g_+)\hat{q}_2 -(g_- +g_+)\hat{q}_2\cos(2\omega_{\rm m} t) -(g_- +g_+)\hat{p}_2\sin(2\omega_{\rm m} t)+\sqrt{2\kappa}\hat{p}_{1,\inpt}, \\
\dot{\hat{q}}_2 &=& -\gamma\hat{q}_2 +(g_- -g_+)\hat{p}_1 -(g_- -g_+)\hat{p}_1\cos(2\omega_{\rm m} t) +(g_- +g_+)\hat{q}_1\sin(2\omega_{\rm m} t)+\sqrt{2\gamma}\hat{q}_{2,\inpt}, \\
\dot{\hat{p}}_2 &=& -\gamma\hat{p}_2 -(g_- +g_+)\hat{q}_1 -(g_- +g_+)\hat{q}_1\cos(2\omega_{\rm m} t) -(g_- -g_+)\hat{p}_1\sin(2\omega_{\rm m} t)+\sqrt{2\gamma}\hat{p}_{2,\inpt}.
\end{eqnarray}
\end{subequations}
Moving to the Floquet basis, we obtain the drift matrix 
\begin{equation}\label{eq:drift2tone}
\mathbf{A}_\Floquet = 
\begin{pmatrix}
\mathbf{A}^{(0)} &\mathbf{A}_{\rm c}^{(1)} &\mathbf{A}_{\rm s}^{(1)} & 0 & 0 \\
\mathbf{A}_{\rm c}^{(1)} &\mathbf{A}^{(0)} & -2\omega_{\rm m}\mathbf{I}_2 & \frac{1}{\sqrt{2}}\mathbf{A}_{\rm c}^{(1)} & \frac{1}{\sqrt{2}}\mathbf{A}_{\rm s}^{(1)} \\
\mathbf{A}_{\rm s}^{(1)} &2\omega_{\rm m}\mathbf{I}_2 &\mathbf{A}^{(0)} & -\frac{1}{\sqrt{2}}\mathbf{A}_{\rm s}^{(1)} & \frac{1}{\sqrt{2}}\mathbf{A}_{\rm c}^{(1)}\\
0 & \frac{1}{\sqrt{2}}\mathbf{A}_{\rm c}^{(1)} & -\frac{1}{\sqrt{2}}\mathbf{A}_{\rm s}^{(1)} & \mathbf{A}^{(0)} & -4\omega_{\rm m}\mathbf{I}_2 \\
0 & \frac{1}{\sqrt{2}}\mathbf{A}_{\rm s}^{(1)} & \frac{1}{\sqrt{2}}\mathbf{A}_{\rm c}^{(1)} & 4\omega_{\rm m}\mathbf{I}_2 & \mathbf{A}^{(0)} \\
\end{pmatrix}\,,
\end{equation}
where
\begin{equation}
	\mathbf{A}^{(0)} = \mathbf{A}_0 + g_-\mathbf{J}_- - g_+\mathbf{J}_+, \qquad
	\mathbf{A}_{\rm c}^{(1)} = -\frac{g_-}{\sqrt{2}}\mathbf{J}_+ + \frac{g_+}{\sqrt{2}}\mathbf{J}_-,\qquad 
	\mathbf{A}_{\rm s}^{(1)} = \frac{g_-}{\sqrt{2}}\mathbf{G}_1 - \frac{g_+}{\sqrt{2}}\mathbf{G}_2,
\end{equation}
where we introduced the matrix
\begin{equation}
	\mathbf{G}_2 = \begin{pmatrix}
		0&0&1&0 \\ 0&0&0&1 \\ -1&0&0&0 \\ 0&-1&0&0
	\end{pmatrix};
\end{equation}
the diffusion matrix (for a single Brillouin zone) is the same as in the case of optomechanical cooling.
Under the \RWA, where the time-dependent parts in Eq.~(\ref{eq:HamDS}) are ignored, we are left with just the Brillouin zone $n=0$, \emph{i.e}, the matrix $\mathbf{A}^{(0)}$.

\begin{figure}
  \centering
  \includegraphics[width=1.0\linewidth]{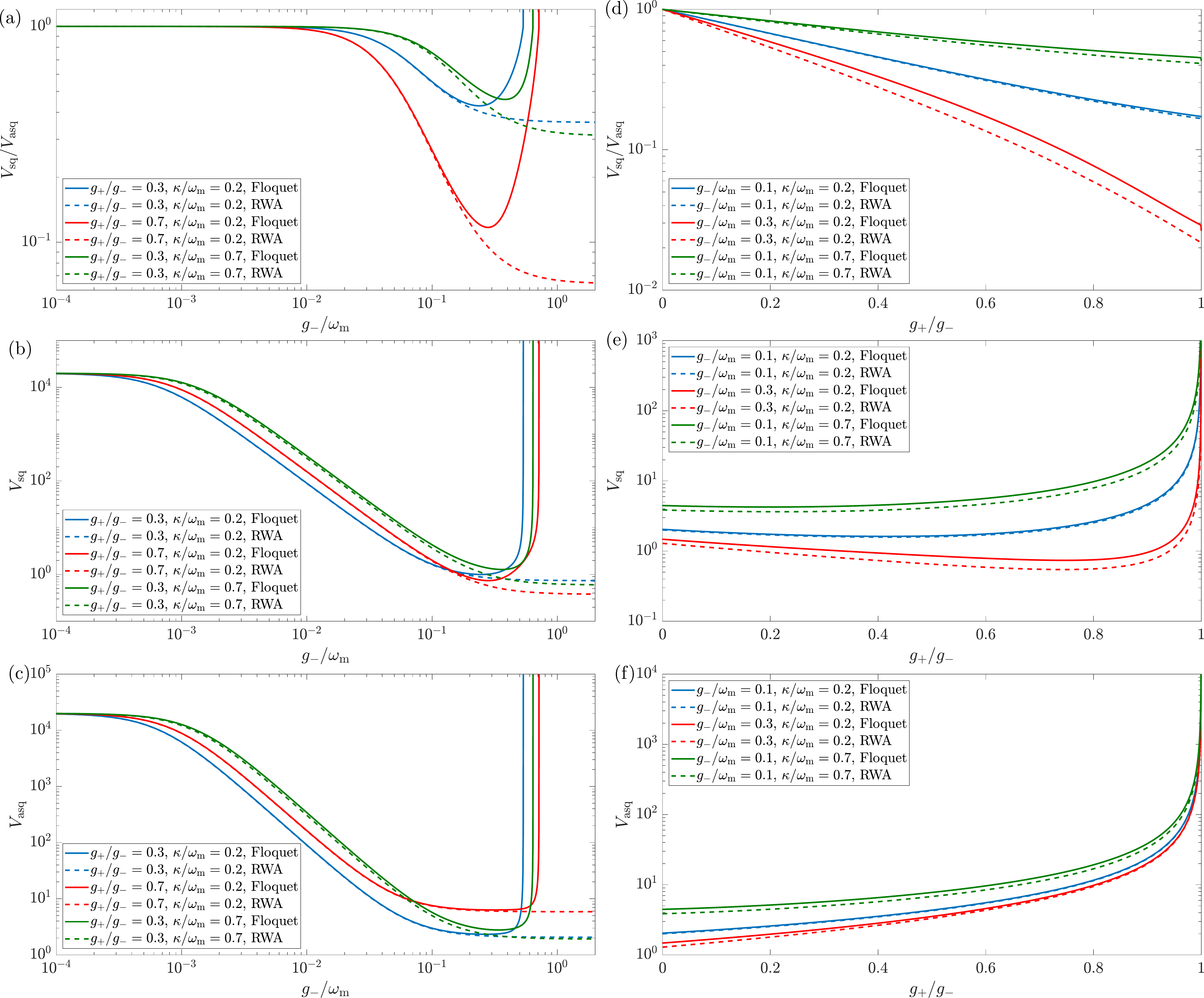}
  \caption{Dissipative mechanical squeezing as a function of (a)--(c) the beam-splitter coupling rate $g_-$ and (d)--(f) the ratio of the two coupling rates $g_+/g_-$.
  	The \RWA{} results are plotted as the dashed lines and the full Floquet--Lyapunov approach with the solid lines.
  	The ratio between the squeezed and antisqueezed variances is plotted in panels (a) and (d),
  	the squeezed variance in panels (b) and (e), and the antisqueezed variance in (c) and (f).
  	The parameters used for the simulations are $\gamma/\omega_{\rm m} = 2\times 10^{-6}$ and $\bar{n}=10^{4}$.}\label{fig:MS}
\end{figure}

To evaluate the mechanical squeezing we need the submatrix describing the mechanical covariances. The solution of the Lyapunov equation is of the form
\begin{equation}
\mathbf{\Gamma}^{(0)} = 
\begin{pmatrix}
\mathbf{\Gamma}_{\rm cav} &\mathbf{\Gamma}_{\rm C} \\
\mathbf{\Gamma}_{\rm C}^\top &\mathbf{\Gamma}_{\rm m}
\end{pmatrix} \,,
\end{equation}
where $\mathbf{\Gamma}^{(0)}$ is a $4\times 4$ matrix. The steady-state squeezed and antisqueezed mechanical variances are
\begin{equation}
V_{\rm sq} = {\rm min}\; {\rm eig}(\mathbf{\Gamma}_{\rm m}), \label{eq:Vsq}\qquad 
V_{\rm asq} = {\rm max}\; {\rm eig}(\mathbf{\Gamma}_{\rm m})  \,.
\end{equation}

The variances simulated by the Floquet--Lyapunov method and under the RWA are shown in Fig.~\ref{fig:MS}.
For small coupling ($g_\pm \ll \omega_{\rm m}$) and a sideband resolved system ($\kappa\ll\omega_{\rm m}$), there is a good agreement between the two approaches.
As the coupling or the cavity decay is increased, particularly the squeezed variance is affected by the presence of the counterrotating terms.
This result is not surprising---as the counterrotating terms give rise to a phase-independent backaction limit, they add a fixed amount of noise to both mechanical quadratures.
As the squeezed variance is smaller, the same amount of added noise corresponds to a larger relative increase of the variance than for the antisqueezed quadrature.
Similar to the case of mechanical cooling in Fig.~\ref{fig:OC1}, the Floquet solution becomes unstable for strong coupling, behaviour not visible with the RWA.

\subsection{Parametric and dissipative squeezing for a levitated particle}\label{ssec:levitation}

In the previous example, only one frequency component (oscillating at $\omega = 2\omega_\mathrm{m}$) was present in the system.
Now, we turn our attention to a system where also the next component is relevant---a levitated particle squeezed by a combination of parametric and dissipative squeezing~\cite{Cernotik2020}.
The potential for the particle's centre-of-mass motion is defined by the laser beam used for levitation;
its scattering into an empty cavity mode provides the optomechanical interaction~\cite{Gonzalez-Ballestero2019,Windey2019,Delic2019}.
To achieve strong squeezing, the optical tweezer amplitude is modulated at twice the mechanical frequency, $E_{\rm tw}(t) = E_0[1+\alpha\cos(2\omega_\mathrm{m} t)]$, where $\alpha\in(0,1)$ is the modulation depth, modulating both the mechanical potential and the optomechanical coupling rate.
Since the potential is proportional to the tweezer intensity (\emph{i.e.}, the square of the amplitude) and the optomechanical coupling to the tweezer amplitude, the system dynamics are characterised by the Hamiltonian~\cite{Cernotik2020}
\begin{equation}
\hat{H} = \frac{\omega_{\rm m}}{2}\hat{p}_2^2 +\frac{\omega_{\rm m}}{2}[1+\alpha\cos(2\omega_{\rm m} t)]^2 \hat{q}_2^2 +\Delta\hat{c}^\dagger\hat{c} -g[1+\alpha\cos(2\omega_{\rm m} t)]\hat{q}_2(\hat{c}^\dagger +\hat{c}) \,. \label{eq:HamLP}
\end{equation}

The parametric squeezing is provided by the modulation of the trapping frequency according to the second term in Eq.~\eqref{eq:HamLP}.
To see how the dissipative squeezing arises, we set the detuning to the red mechanical sideband, $\Delta = \omega_\mathrm{m}$, and move to the rotating frame with respect to the free oscillations, $\hat{H}_0 = \omega_\mathrm{m}\hat{c}^\dagger\hat{c}+\frac{1}{2}\omega_\mathrm{m}(\hat{q}_2^2+\hat{p}_2^2)$;
under the \RWA, we then obtain the interaction-picture Hamiltonian
\begin{equation}\label{eq:HamLP}
	\Hint = \frac{\omega_\mathrm{m}\alpha}{4}(\hat{b}^2+\hat{b}^{\dagger 2}+\alpha\hat{b}^\dagger\hat{b}) -\frac{g}{\sqrt{2}}\left(\hat{b}+\frac{\alpha}{2}\hat{b}^\dagger\right)\hat{c}^\dagger -\frac{g}{\sqrt{2}}\left(\hat{b}^\dagger+\frac{\alpha}{2}\hat{b}\right)\hat{c}.
\end{equation}
Introducing the mechanical Bogoliubov mode $\hat{\beta} = (2\hat{b}+\alpha\hat{b}^\dagger)/\sqrt{4-\alpha^2}$ (which satisfies $[\hat{\beta},\hat{\beta}^\dagger] = 1$), we can rewrite this Hamiltonian as
\begin{equation}
	\Hint = -\frac{\omega_{\rm m}\alpha^2}{4}\hat{\beta}^\dagger\hat{\beta} + \frac{\omega_{\rm m}\alpha}{4}(\hat{\beta}^2+\hat{\beta}^{\dagger 2}) - \lambda_{\rm eff}(\hat{c}^\dagger\hat{\beta}+\hat{\beta}^\dagger\hat{c}),
\end{equation}
where $\lambda_{\rm eff} = \lambda\sqrt{(4-\alpha^2)/8}$.
From this expression, we see that the Bogoliubov mode undergoes slow oscillations at frequency $\omega_{\rm m}\alpha^2\ll\omega_{\rm m}$ (the first term on the right-hand side) while being squeezed parametrically (the second term in the Hamiltonian);
dissipative squeezing is provided by the optomechanical interaction in the last term.
Remarkably, these two squeezing techniques can combine and provide stronger squeezing than either strategy alone~\cite{Cernotik2020}.

To describe the generated squeezing beyond the \RWA{}, we write the interaction Hamiltonian~\eqref{eq:HamLP} including the counterrotating terms,
\begin{align}
	\Hint &= \frac{\omega_{\rm m}\alpha}{4}\left(\hat{b}^2 +\hat{b}^{\dagger 2} +\alpha\hat{b}^\dagger\hat{b} \right) \nonumber 
		-\frac{g}{\sqrt{2}}\left(\hat{b}+\frac{\alpha}{2}\hat{b}^\dagger\right)\hat{c}^\dagger
		-\frac{g}{\sqrt{2}}\left(\hat{b}^\dagger+\frac{\alpha}{2}\hat{b}\right)\hat{c} 
		+\left[\omega_{\rm m}\alpha\hat{b}^\dagger\hat{b}+\frac{\omega_{\rm m}\alpha^2}{8}(\hat{b}^2+\hat{b}^{\dagger 2})-\frac{g\alpha}{\sqrt{2}}(\hat{c}^\dagger\hat{b}+\hat{b}^\dagger\hat{c})\right]\cos(2\omega_{\rm m}t)\\
	&\quad +\left(\frac{\omega_{\rm m}\alpha^2}{16}\hat{b}^2 -\frac{g}{\sqrt{2}}\hat{c}\hat{b}\right)\rme^{-2\rmi\omega_{\rm m} t}
		+\left(\frac{\omega_{\rm m}\alpha^2}{16}\hat{b}^{\dagger 2} -\frac{g}{\sqrt{2}}\hat{c}^\dagger\hat{b}^\dagger\right)\rme^{2\rmi\omega_{\rm m} t}
		+\frac{\omega_{\rm m}\alpha^2}{4}\hat{b}^\dagger\hat{b}\cos(4\omega_{\rm m}t) \\
	&\quad +\left(\frac{\omega_{\rm m}\alpha}{4}\hat{b}^2 -\frac{g\alpha}{2\sqrt{2}} \hat{c}\hat{b} \right)\rme^{-4\rmi\omega_{\rm m} t}
		+\left(\frac{\omega_{\rm m}\alpha}{4}\hat{b}^{\dagger 2} -\frac{g\alpha}{2\sqrt{2}} \hat{c}^\dagger\hat{b}^\dagger \right)\rme^{4\rmi\omega_{\rm m} t}
		+\frac{\omega_{\rm m}\alpha^2}{16} (\hat{b}^{\dagger 2} \rme^{6\rmi\omega_{\rm m} t} +\hat{b}^2 \rme^{-6\rmi\omega_{\rm m} t}) \,. \nonumber
\end{align}
From this Hamiltonian, we derive the equations of motion
\begin{subequations}
\begin{align}
	\dot{\hat{q}}_1 &= -\kappa\hat{q}_1 
		-\frac{g}{\sqrt{2}}\left(\sin(2\omega_{\rm m}t) +\frac{\alpha}{2}\sin(4\omega_{\rm m}t)\right)\hat{q}_2
		-\frac{g}{\sqrt{2}}\left(\frac{2-\alpha}{2}-(1-\alpha)\cos(2\omega_{\rm m}t)-\frac{\alpha}{2}\cos(4\omega_{\rm m}t)\right)\hat{p}_2 
		+\sqrt{2\kappa}\,\hat{q}_{1,\inpt}, \\
	\dot{\hat{p}}_1 &= -\kappa\hat{p}_1
		+\frac{g}{\sqrt{2}}\left(\frac{2+\alpha}{2} +(1+\alpha)\cos(2\omega_{\rm m} t) +\frac{\alpha}{2}\cos(4\omega_{\rm m}t)\right)\hat{q}_2
		+\frac{g}{\sqrt{2}}\left(\sin(2\omega_{\rm m} t) +\frac{\alpha}{2}\sin(4\omega_{\rm m}t)\right) \hat{p}_2 +\sqrt{2\kappa}\,\hat{p}_{1,\inpt}, \\
	\begin{split}
	\dot{\hat{q}}_2 &= -\frac{g}{\sqrt{2}}\left(\sin(2\omega_{\rm m}t) +\frac{\alpha}{2}\sin(4\omega_{\rm m}t)\right)\hat{q}_1
		-\frac{g}{\sqrt{2}}\left(\frac{2-\alpha}{2} -(1-\alpha)\cos(2\omega_{\rm m}t) -\frac{\alpha}{2}\cos(4\omega_{\rm m}t)\right)\hat{p}_1 \\
		&\quad  -\gamma\hat{q}_2 +\frac{\omega_{\rm m}\alpha}{2}\left(\frac{\alpha}{4}\sin(2\omega_{\rm m}t) +\sin(4\omega_{\rm m}t) +\frac{\alpha}{4}\sin(6\omega_{\rm m}t)\right)\hat{q}_2 \\
		&\quad -\frac{\omega_{\rm m}\alpha}{4}\left((2-\alpha) -\frac{8-3\alpha}{2}\cos(2\omega_{\rm m}t) +(2-\alpha)\cos(4\omega_{\rm m}t) +\frac{\alpha}{2}\cos(6\omega_{\rm m}t) \right)\hat{p}_2 +\sqrt{2\gamma}\,\hat{q}_{2,\inpt},
	\end{split}\\
	\begin{split}
	\dot{\hat{p}}_2 &= \frac{g}{\sqrt{2}}\left(\frac{2+\alpha}{2} +(1+\alpha)\cos(2\omega_{\rm m}t) +\frac{\alpha}{2}\cos(4\omega_{\rm m}t) \right)\hat{q}_1
		+\frac{g}{\sqrt{2}}\left(\sin(2\omega_{\rm m}t) +\frac{\alpha}{2}\sin(4\omega_{\rm m}t) \right)\hat{p}_1 \\
		&\quad -\frac{\omega_{\rm m}\alpha}{4}\left((2+\alpha) +\frac{8+3\alpha}{2}\cos(2\omega_{\rm m}t) +(2+\alpha)\cos(4\omega_{\rm m}t) +\frac{\alpha}{2}\cos(6\omega_{\rm m}t) \right)\hat{q}_2  \\
		&\quad -\gamma\hat{p}_2 -\frac{\omega_{\rm m}\alpha}{2}\left(\frac{\alpha}{4}\sin(2\omega_{\rm m}t) +\sin(4\omega_{\rm m} t) +\frac{\alpha}{4}\sin(6\omega_{\rm m}t) \right)\hat{p}_2 +\sqrt{2\gamma}\,\hat{p}_{2,\inpt},
	\end{split}
\end{align}
\end{subequations}
from which we obtain the drift matrix
\begin{equation}\label{eq:driftLev}
	\mathbf{A}_\Floquet = 
	\begin{pmatrix}
		\mathbf{A}^{(0)} &\mathbf{A}_{\rm c}^{(1)} &\mathbf{A}_{\rm s}^{(1)} &\mathbf{A}_{\rm c}^{(2)} &\mathbf{A}_{\rm s}^{(2)} &
			\mathbf{A}_{\rm c}^{(3)} & \mathbf{A}_{\rm s}^{(3)}\\
		\mathbf{A}_{\rm c}^{(1)} &\mathbf{A}^{(0)} +\frac{1}{\sqrt{2}}\mathbf{A}_{\rm c}^{(2)} &
			-2\omega_{\rm m}\mathbf{I}_2 +\frac{1}{\sqrt{2}}\mathbf{A}_{\rm s}^{(2)} &\frac{1}{\sqrt{2}}(\mathbf{A}_{\rm c}^{(1)} +\mathbf{A}_{\rm c}^{(3)}) &
			\frac{1}{\sqrt{2}}(\mathbf{A}_{\rm s}^{(1)} +\mathbf{A}_{\rm s}^{(3)}) & 
			\frac{1}{\sqrt{2}}\mathbf{A}_{\rm c}^{(2)} & \frac{1}{\sqrt{2}}\mathbf{A}_{\rm s}^{(2)}\\
		\mathbf{A}_{\rm s}^{(1)} &2\omega_{\rm m}\mathbf{I}_2 +\frac{1}{\sqrt{2}}\mathbf{A}_{\rm s}^{(2)} &
			\mathbf{A}^{(0)} -\frac{1}{\sqrt{2}}\mathbf{A}_{\rm c}^{(2)} &\frac{1}{\sqrt{2}}(-\mathbf{A}_{\rm s}^{(1)} +\mathbf{A}_{\rm s}^{(3)}) &
			\frac{1}{\sqrt{2}}(\mathbf{A}_{\rm c}^{(1)} -\mathbf{A}_{\rm c}^{(3)}) & -\frac{1}{\sqrt{2}}\mathbf{A}_{\rm s}^{(2)}
			& \frac{1}{\sqrt{2}}\mathbf{A}_{\rm c}^{(2)} \\
		\mathbf{A}_{\rm c}^{(2)} &\frac{1}{\sqrt{2}}(\mathbf{A}_{\rm c}^{(1)} +\mathbf{A}_{\rm c}^{(3)}) &
			\frac{1}{\sqrt{2}}(-\mathbf{A}_{\rm s}^{(1)} +\mathbf{A}_{\rm s}^{(3)}) &\mathbf{A}^{(0)} &-4\omega_{\rm m}\mathbf{I}_2 & 
			\frac{1}{\sqrt{2}}\mathbf{A}_{\rm c}^{(1)} & \frac{1}{\sqrt{2}}\mathbf{A}_{\rm s}^{(1)}  \\
		\mathbf{A}_{\rm s}^{(2)} &\frac{1}{\sqrt{2}}(\mathbf{A}_{\rm s}^{(1)} +\mathbf{A}_{\rm s}^{(3)}) &
			\frac{1}{\sqrt{2}}(\mathbf{A}_{\rm c}^{(1)} -\mathbf{A}_{\rm c}^{(3)}) &4\omega_{\rm m}\mathbf{I}_2 &\mathbf{A}^{(0)} & 
			-\frac{1}{\sqrt{2}}\mathbf{A}_{\rm s}^{(1)} & \frac{1}{\sqrt{2}}\mathbf{A}_{\rm c}^{(1)} \\
		\mathbf{A}_{\rm c}^{(3)} & \frac{1}{\sqrt{2}}\mathbf{A}_{\rm c}^{(2)} & -\frac{1}{\sqrt{2}}\mathbf{A}_{\rm s}^{(2)} &
			\frac{1}{\sqrt{2}}\mathbf{A}_{\rm c}^{(1)} & -\frac{1}{\sqrt{2}}\mathbf{A}_{\rm s}^{(1)} & \mathbf{A}^{(0)} & -6\omega_{\rm m}\mathbf{I}_2 \\
		\mathbf{A}_{\rm s}^{(3)} & \frac{1}{\sqrt{2}}\mathbf{A}_{\rm s}^{(2)} & \frac{1}{\sqrt{2}}\mathbf{A}_{\rm c}^{(2)} &
			\frac{1}{\sqrt{2}}\mathbf{A}_{\rm s}^{(1)} & \frac{1}{\sqrt{2}}\mathbf{A}_{\rm c}^{(1)} &6\omega_{\rm m}\mathbf{I}_2 & \mathbf{A}^{(0)}
	\end{pmatrix}\,,
\end{equation}
where
\begin{subequations}
\begin{align}
	\mathbf{A}^{(0)} &= \mathbf{A}_0 -\frac{g}{\sqrt{2}}\mathbf{J}_- +\frac{g\alpha}{2\sqrt{2}}\mathbf{J}_+ +\frac{\omega_{\rm m}\alpha^2}{4}\mathbf{M}_- -\frac{\omega_{\rm m}\alpha}{2}\mathbf{M}_+, \\
	\mathbf{A}_{\rm c}^{(1)} &= \frac{g}{2}\mathbf{J}_+ -\frac{g\alpha}{2}\mathbf{J}_- +\frac{\omega_{\rm m}\alpha}{\sqrt{2}}\mathbf{M}_- -\frac{3\omega_{\rm m}\alpha^2}{8\sqrt{2}}\mathbf{M}_+, \qquad 
	\mathbf{A}_{\rm s}^{(1)} = -\frac{g}{2}\mathbf{G}_1 + \frac{\omega_{\rm m}\alpha^2}{8\sqrt{2}}\mathbf{M}_0, \\
	\mathbf{A}_{\rm c}^{(2)} &= \frac{g\alpha}{4}\mathbf{J}_+ -\frac{\omega_{\rm m}\alpha}{2\sqrt{2}}\mathbf{M}_+ +\frac{\omega_{\rm m}\alpha^2}{4\sqrt{2}}\mathbf{M}_-, \qquad 
	\mathbf{A}_{\rm s}^{(2)} = -\frac{g\alpha}{4}\mathbf{G}_1 + \frac{\omega_{\rm m}\alpha}{2\sqrt{2}}\mathbf{M}_0, \\
	\mathbf{A}_{\rm c}^{(3)} &= -\frac{\omega_{\rm m} \alpha^2}{8\sqrt{2}}\mathbf{M}_+, \qquad 
	\mathbf{A}_{\rm s}^{(3)} = \frac{\omega_{\rm m} \alpha^2}{8\sqrt{2}} \mathbf{M}_0 
\end{align}
\end{subequations}
with
\begin{equation}
	\mathbf{M}_\pm = \begin{pmatrix}
		0&0&0&0 \\ 0&0&0&0 \\ 0&0&0&1 \\ 0&0&\pm 1&0
	\end{pmatrix},\qquad
	\mathbf{M}_0 = {\rm diag}(0,0,1,-1).
\end{equation}
Again, the diffusion matrix is the same as before and the \RWA{} corresponds to working with the zeroth Brillouin zone only.

\begin{figure}
  \centering
  \includegraphics[width=1.0\linewidth]{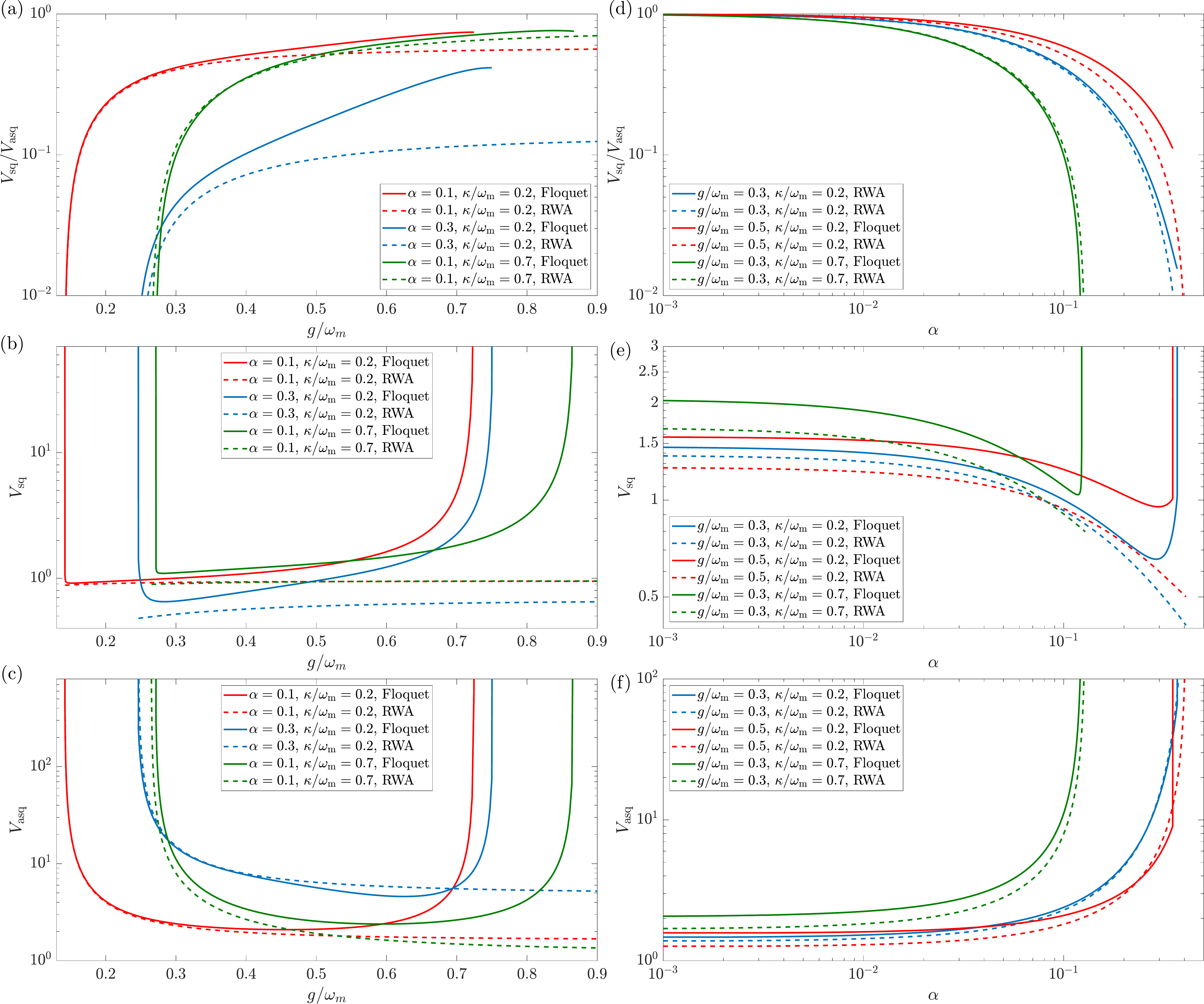}
  \caption{Parametric and dissipative mechanical squeezing for a levitated particle as a function of (a)--(c) the coupling strength $g$ and (d)--(f) modulation depth $\alpha$.
  	The lines have the same meaning as in Fig.~\ref{fig:MS} and we again plot the ratio of the variances (top), the squeezed variance (middle) and the anti-squeezed variance (bottom).
  	The parameters used for the simulations are $\gamma/\omega_{\rm m} = 10^{-9}$ and $\bar{n}=2\times 10^{7}$.}\label{fig:LP}
\end{figure}

We compare the results of the Floquet--Lyapunov approach and the \RWA{} in Fig.~\ref{fig:LP}.
Again, the squeezed quadrature is more sensitive to the validity of the \RWA{} than the antisqueezed quadrature;
generally, this effect seems more pronounced now owing to the larger number of counterrotating terms present in the system.
Surprisingly, there exist regimes for the levitated particle where the antisqueezed variance is smaller with the counterrotating terms than under the \RWA{} (see the red lines in Fig.~\ref{fig:LP}(f)).
This result points to a nontrivial role that the counterrotating terms play in this situation, which might be harnessed by optimising the modulation phase in the tweezer amplitude, $E_{\rm tw}(t) = E_0[1+\alpha\cos(2\omega_{\rm m}t+\phi)]$ (we considered $\phi = 0$ for simplicity here).

The dynamical stability of the system is now more complicated than in the case of dissipative squeezing via a two-tone drive.
Similar to the previous case, the system becomes unstable for large coupling, which is not captured by the \RWA{} calculation.
Additionally, the system is unstable (both under \RWA{} and with the Floquet--Lyapunov approach) also for large modulation depth, where the parametric squeezing effect becomes too strong and for weak optomechanical coupling.
We can understand the latter result by noting that, for $g\to 0$, only the parametric squeezing effect persists;
for the modulation depth considered in Fig.~\ref{fig:LP}, this would correspond to more than 3 dB of squeezing, making the system unstable.

\section{Conclusions}

In this manuscript, we presented an efficient method for analysing dynamics of Gaussian systems with periodic Hamiltonians.
Our approach is based on deriving equations of motion for the canonical operators in the Floquet space from which a Lyapunov equation for the covariance matrix of the system's Wigner function can be formulated.
In this way, we arrive at a time-independent algebraic equation for the steady state of the system starting from a time-dependent one;
the price we pay for this---increased size of the Hilbert space---is not prohibitive for numerical simulations.
We expect the technique to find wide use in cavity optomechanics and electromechanics where time-dependent coupling rates are commonly used for dissipative state preparation.
In these systems, large thermal phonon occupations of the mechanical reservoirs preclude direct simulations of the master equation in the Fock basis.

We demonstrated the use of our techniques on dissipative generation of mechanical squeezing in conventional dispersive and novel levitated optomechanical systems and performed a detailed analysis of the role counterrotating terms play in these systems.
We showed the intuitive result that the squeezed quadrature is more sensitive to the validity of the \RWA{} than the antisqueezed quadrature owing to its overall lower amount of noise.
Importantly, even though counterrotating terms have stronger effect in the recent proposal of combined parametric and dissipative squeezing for levitated particles~\cite{Cernotik2020}, sub-vacuum squeezing is still possible with state-of-the-art systems;
our simulations show that about 2 dB of squeezing (with about 11 dB of antisqueezing) can be generated with parameters similar to Ref.~\cite{Delic2020}.
Finally, we note that the presence of counterrotating terms can also affect the dynamical stability of the system (particularly when approaching the ultrastrong coupling regime) which cannot be seen from simple analysis under the \RWA{}.

Our approach can be readily adapted for various other state preparation schemes that rely on multitone driving, such as generation of Gaussian entanglement between two mechanical resonators~\cite{Tan2013,Woolley2014}.
Our technique allows a straightforward analysis of dissipative state preparation beyond the \RWA{} and a direct evaluation of approaches to mitigate the undesired effects of the counterrotating terms.
One such possible technique is using cavity fields with squeezed input noise.
This strategy has already been used to suppress the heating associated with counterrotating terms in sideband cooling~\cite{Asjad2016,Clark2017};
it would be interesting to see whether its benefits persist also in more complex dissipative dynamics such as generation of one- and two-mode mechanical squeezing.

Our strategy could also be used to analyse conditional dynamics of optomechanical systems under backaction-evading measurements~\cite{Clerk2008,Brunelli2019}.
In these schemes, driving on both mechanical sidebands is used to perform a quantum nondemolition measurement of one of the mechanical quadratures;
similar to dissipative state preparation, the effects of counterrotating terms are often neglected under the \RWA{}.
Since the measurement at the output is homodyne (projecting the system onto a Gaussian state), the system can still be fully characterised using the covariance matrix formalism;
the important distinction is that the dynamics of the conditional state obey an algebraic equation of the Riccati type~\cite{Edwards2005}.
The Riccati equation is closely related to the Lyapunov equation---it operates with the same drift and diffusion matrices (which account for the unconditional part of the evolution) with additional, nonlinear terms stemming from the weak measurement.
Since the measurement is typically time-independent in backaction-evading measurements, extension of our formalism to Riccati-type equations should be straightforward.

All in all, dissipative preparation of nonclassical mechanical states and their tomography using quantum nondemolition measurements are important techniques for modern optomechanical and electromechanical systems.
As we have discussed, they are often based on the use of periodic Hamiltonians in which resonant parts give rise to the desired dynamics and fast oscillating terms are neglected under the \RWA{}.
The \RWA{} can be invalidated by two factors: nonzero sideband ratio (between the cavity damping and mechanical frequency), which stems from practical limitations in realistic experimental systems, and ultrastrong coupling, associated with improved experimental design.
With our techniques, the dynamics of these systems can be efficiently analysed including the effects of the fast oscillating terms, allowing the new physics associated with these processes to be uncovered.

\begin{acknowledgments}
	Our work was supported by the project LTAUSA19099 from the Czech Ministry of Education, Youth and Sports (MEYS \v{C}R),
	European Union’s Horizon 2020 (2014–2020) research and innovation framework programme under Grant Agreement No. 731473 (project 8C18003 TheBlinQC) (I.P. and R.F.),
	and the project CZ.02.1.01/0.0/0.0/16\textunderscore{}026/0008460 of MEYS \v{C}R (O.\v{C}. and R.F.).
	R.F. was also supported by the Project 20-16577S of the Czech Science Foundation.
	Project TheBlinQC has received funding from the QuantERA ERA-NET Cofund in Quantum Technologies implemented within the European Union’s Horizon 2020 Programme.
\end{acknowledgments}

\appendix

\section{Floquet method}\label{app:Floquet}

According to the Floquet theory, the solution to a linear system $\dot{\mathbf{y}} = \mathbf{B}(t)\mathbf{y}$, where $ \mathbf{B}(t)$ is periodic, can be written in the form $\dot{\mathbf{y}}_\Floquet = \mathbf{B}_\Floquet \mathbf{y}_\Floquet$, where $\mathbf{B}_\Floquet$ is time independent. 
The vectors transform as $\mathbf{y} = \mathbf{Q}(t) \mathbf{y}_\Floquet$, where $\mathbf{Q}(t)$ is periodic;
we will show in this section that the time-independent matrix $\mathbf{B}_\Floquet$ can be found from
\begin{equation}\label{eqA:QA}
	\mathbf{QB}_\Floquet = \mathbf{BQ} -\dot{\mathbf{Q}}\,.
\end{equation}

Let us consider a time-dependent Hamiltonian that is $\tau$-periodic, $\hat{H}(t) = \hat{H}(t+\tau)$, and quadratic in the canonical operators. Representing all modes with creation ($\hat{a}_i^\dagger$) and annihilation ($\hat{a}_i$) operators, we can write the time development of their expectation values $\alpha_i = \avg{\hat{a}_i}, \alpha_i^\ast = \avg{\hat{a}_i^\dagger}$ as
\begin{equation}\label{eqA:LinSys}
\dot{\mathbf{y}} = \mathbf{B}(t)\mathbf{y} \,,
\end{equation}
where we defined the vector $\mathbf{y} = (\alpha_1, \alpha_1^\ast, \ldots, \alpha_N, \alpha_N^\ast )^\top \,$.
Since the linear system~\eqref{eqA:LinSys} is $\tau$-periodic with the corresponding frequency $\omega = 2\pi/\tau$, we can decompose $\mathbf{y}$ and $\mathbf{B}(t)$ in their frequency components according to
\begin{align}\label{eqA:Fourier}
	\alpha_i = \sum_{n=-\infty}^\infty \alpha_i^{(n)} \rme^{\rmi n\omega t} \,,\qquad 
	\alpha_i^\ast = \sum_{n=-\infty}^\infty \alpha_i^{\ast (n)} \rme^{\rmi n\omega t}  \,, \qquad 
	\mathbf{B}(t) = \sum_{n=-\infty}^\infty \mathbf{B}^{(n)} \rme^{\rmi n\omega t} \,.
\end{align}	
Defining the vector of frequency components $\mathbf{y}^{(n)} = (\alpha_1^{(n)},\alpha_1^{(n)\ast},\ldots,\alpha_N^{(n)},\alpha_N^{(n)\ast})^\top$ and the Floquet-space vector $\mathbf{y}_\Floquet = (\ldots,\mathbf{y}^{(n)},\mathbf{y}^{(n+1)},\ldots)^\top$,
we can write the relation $\mathbf{y} = \mathbf{Q}(t)\mathbf{y}_\Floquet$ with the $2N\times\infty$ matrix
\begin{equation}
	\mathbf{Q}(t) = 
	\begin{pmatrix}
		\ldots,\rme^{\rmi n\omega t}\mathbf{I}_N,\rme^{\rmi(n+1)\omega t}\mathbf{I}_N,\ldots
	\end{pmatrix},
\end{equation}
where $\mathbf{I}_N$ is the $N$-mode (and hence $2N$-dimensional) identity matrix.
With the frequency decomposition of $\mathbf{B}(t)$ and the matrix $\mathbf{Q}(t)$, we can evaluate Eq.~(\ref{eqA:QA});
a straightforward calculation reveals that individual frequency components obey the differential equation
\begin{equation}\label{eqA:EoMx}
\dot{\mathbf{y}}^{(n)} = \sum_m (\mathbf{B}^{(m)} -\rmi n\omega\delta_{m0})\mathbf{y}^{(n-m)} .
\end{equation}

Transformation from the frequency components of the creation and annihilation operators (defined in terms of the complex exponential $e^{in\omega t}$) to the sine and cosine components of the quadratures (as introduced in Eqs.~\eqref{eq:Fourier}) is possible by introducing the definitions
\begin{subequations}
\begin{align}
	q_i^{(0)} = \frac{\alpha_i^{(0)}+\alpha_i^{\ast (0)}}{\sqrt{2}}, \qquad &
	p_i^{(0)} = -\rmi\frac{\alpha_i^{(0)}-\alpha_i^{\ast (0)}}{\sqrt{2}}, \\
	q_{i,\mathrm{c}}^{(n)} = \frac{\alpha_i^{(n)} +\alpha_i^{(-n)} +\alpha_i^{\ast (n)} +\alpha_i^{\ast (-n)}}{2}, \qquad &
	p_{i,\mathrm{c}}^{(n)} = -\rmi\frac{\alpha_i^{(n)} +\alpha_i^{(-n)} -\alpha_i^{\ast (n)} -\alpha_i^{\ast (-n)}}{2}, \\
	q_{i,\mathrm{s}}^{(n)} = \rmi\frac{\alpha_i^{(n)} -\alpha_i^{(-n)} +\alpha_i^{\ast (n)} -\alpha_i^{\ast (-n)}}{2}, \qquad &
	p_{i,\mathrm{s}}^{(n)} = \frac{\alpha_i^{(n)} -\alpha_i^{(-n)} -\alpha_i^{\ast (n)} +\alpha_i^{\ast (-n)}}{2}.
\end{align}
\end{subequations}
Note that these quantities are real since, owing to the definitions of the frequency components in Eq.~\eqref{eqA:Fourier}, we have $(\alpha_i^{(n)})^\ast = \alpha_i^{\ast (-n)}$.
Moreover, the normalisation of the operators is chosen such that the basis change is unitary;
this choice also fixes the prefactor in front of the sum in the Fourier transform
\begin{equation}
	f(t) = f^{(0)} + \sqrt{2}\sum_{n=1}^\infty [f_\mathrm{c}^{(n)}\cos(n\omega t) + f_\mathrm{s}^{(n)}\sin(n\omega t)].
\end{equation}
Indeed, introducing the vectors $\mathbf{r}^{(0)} = (q_1^{(0)},p_1^{(0)},\ldots,q_N^{(0)},p_N^{(0)})^\top$, $\mathbf{r}_\mathrm{c}^{(n)} = (q_{1,\mathrm{c}}^{(n)},p_{1,\mathrm{c}}^{(n)},\ldots,q_{N,\mathrm{c}}^{(n)},p_{N,\mathrm{c}}^{(n)})^\top$, and $\mathbf{r}_\mathrm{s}^{(n)} = (q_{1,\mathrm{s}}^{(n)},p_{1,\mathrm{s}}^{(n)},\ldots,q_{N,\mathrm{s}}^{(n)},p_{N,\mathrm{s}}^{(n)})^\top$, as well as the Floquet vector $\mathbf{r}_\Floquet = (\mathbf{r}^{(0)},\mathbf{r}_\mathrm{c}^{(1)},\mathbf{r}_\mathrm{s}^{(1)},\ldots)^\top$, we can write
\begin{equation}
	\mathbf{r}_\Floquet = \mathbf{Ry}_\Floquet,
\end{equation}
where we introduced the unitary matrix
\begin{subequations}
\begin{align}
	\mathbf{R} &= 
	\begin{pmatrix}
		& 0 & 0 & \mathbf{R}_0 & 0 & 0 \\
		& 0 & \mathbf{R}_\mathrm{c} & 0 & \mathbf{R}_\mathrm{c}  & 0 \\
		& 0 & -\mathbf{R}_\mathrm{s} & 0 & \mathbf{R}_\mathrm{s} & 0 \\
		& \mathbf{R}_\mathrm{c} & 0 & 0 & 0 & \mathbf{R}_\mathrm{c} \\
		& -\mathbf{R}_\mathrm{s} & 0 & 0 & 0 & \mathbf{R}_\mathrm{s} \\
		\udots &&&\vdots &&& \ddots 
	\end{pmatrix},\\
	\mathbf{R}_0 &= \frac{1}{\sqrt{2}}\begin{pmatrix}
		1 & 1 \\
		-\rmi & \rmi \\
		&& 1 & 1\\
		&& -\rmi & \rmi\\
		&&&&\ddots 
	\end{pmatrix}, \qquad 
	\mathbf{R}_\mathrm{c} = \frac{1}{2}\begin{pmatrix}
		1&1 \\
		-\rmi &\rmi \\
		&& 1&1 \\
		&& -\rmi &\rmi \\
		&&&& \ddots 
	\end{pmatrix} = \frac{1}{\sqrt{2}}\mathbf{R}_0, \qquad 
	\mathbf{R}_\mathrm{s} = \frac{1}{2}\begin{pmatrix}
		\rmi &\rmi \\
		1&-1 \\
		&& \rmi &\rmi \\
		&& 1&-1 \\
		&&&&\ddots 
	\end{pmatrix} = \frac{\rmi}{\sqrt{2}}\mathbf{R}_0.
\end{align}
\end{subequations}
We can now transform the Langevin equation in the Floquet space, $\dot{\mathbf{y}}_\Floquet = \mathbf{B}_\Floquet\mathbf{y}_\Floquet$, into
\begin{equation}
	\dot{\mathbf{r}}_\Floquet = \mathbf{R}\dot{\mathbf{y}}_\Floquet = \mathbf{RA}_\Floquet\mathbf{R}^\dagger \mathbf{R y}_\Floquet = \mathbf{A}_\Floquet\mathbf{r}_\Floquet,
\end{equation}
where $\mathbf{A}_\Floquet$ is given by Eq.~\eqref{eq:Ar} in the main text.

So far we assumed that the system is homogeneous, which is valid only for the classical expectation values $\alpha_i = \avg{\hat{a}_i}$, $q_i = \avg{\hat{q}_i}$, $p_i = \avg{\hat{p}_i}$.
To include quantum fluctuations, we append the original linear system~\eqref{eqA:LinSys} with a vector of noise operators $\hat{\xi}_y = (\hat{a}_{1,\inpt},\hat{a}_{1,\inpt}^\dagger,\ldots,\hat{a}_{N,\inpt},\hat{a}_{N,\inpt}^\dagger)^\top$,
\begin{equation}
	\mathbf{\dot{\hat{y}}} = \mathbf{B}(t)\mathbf{\hat{y}}+\hat{\xi}_y,
\end{equation}
where we now work with the vector of operators $\mathbf{\hat{y}} = (\hat{a}_1,\hat{a}_1^\dagger,\ldots,\hat{a}_N,\hat{a}_N^\dagger)^\top$ instead of their expectation values $\mathbf{y} = \avg{\mathbf{\hat{y}}}$.
Since the system is Markovian, the noise operators have flat spectra and can thus be separated into their individual frequency components without changing their correlation properties.
This allows us to proceed in full analogy to the classical expectation values and write a quantum version of Eq.~\eqref{eqA:EoMx} in the form
\begin{equation}
	\mathbf{\dot{\hat{y}}}^{(n)} = \sum_m(\mathbf{B}^{(m)}-\rmi n\omega\delta_{m0})\mathbf{\hat{y}}^{(n-m)}+\hat{\xi}_y^{(n)}.
\end{equation}
Transformation from annihilation and creation operators to canonical operators then gives the time-independent Langevin equation
\begin{equation}
	\mathbf{\dot{\hat{r}}}_\Floquet = \mathbf{A}_\Floquet \mathbf{\hat{r}}_\Floquet + \hat{\xi}_\Floquet
\end{equation}
with $\hat{\xi}_\Floquet$ defined in full analogy with $\mathbf{\hat{r}}_\Floquet$.
Crucially, Markovianity ensures that the individual frequency components of the noise are uncorrelated so we can write
\begin{equation}
	\mathbf{N}^{(0)} = \avg{\hat{\xi}^{(0)}\hat{\xi}^{(0)\top}}= \mathbf{N}_{\rm k}^{(n)} = \avg{\hat{\xi}_{\rm k}^{(n)}\hat{\xi}_{\rm k}^{(n)\top}} = \avg{\hat{\xi}\hat{\xi}^{\top}} = \mathbf{N},
\end{equation}
where ${\rm k}\in\{\mathrm{c,s}\}$ and $\hat{\xi}$ is the input quadrature noise of the original system, and $\mathbf{N}_\Floquet = \mathrm{diag}(\ldots,\mathbf{N},\mathbf{N},\ldots)$.
The Floquet-space covariance matrix $\mathbf{\Gamma}_\Floquet$ then obeys the Lyapunov equation
\begin{equation}
	\dot{\mathbf{\Gamma}}_\Floquet = \mathbf{A}_\Floquet \mathbf{\Gamma}_\Floquet + \mathbf{\Gamma}_\Floquet \mathbf{A}_\Floquet^\top +\mathbf{N}_\Floquet.
\end{equation}

\section{Convergence}

\begin{figure}
  \centering
  \includegraphics[width=0.8\linewidth]{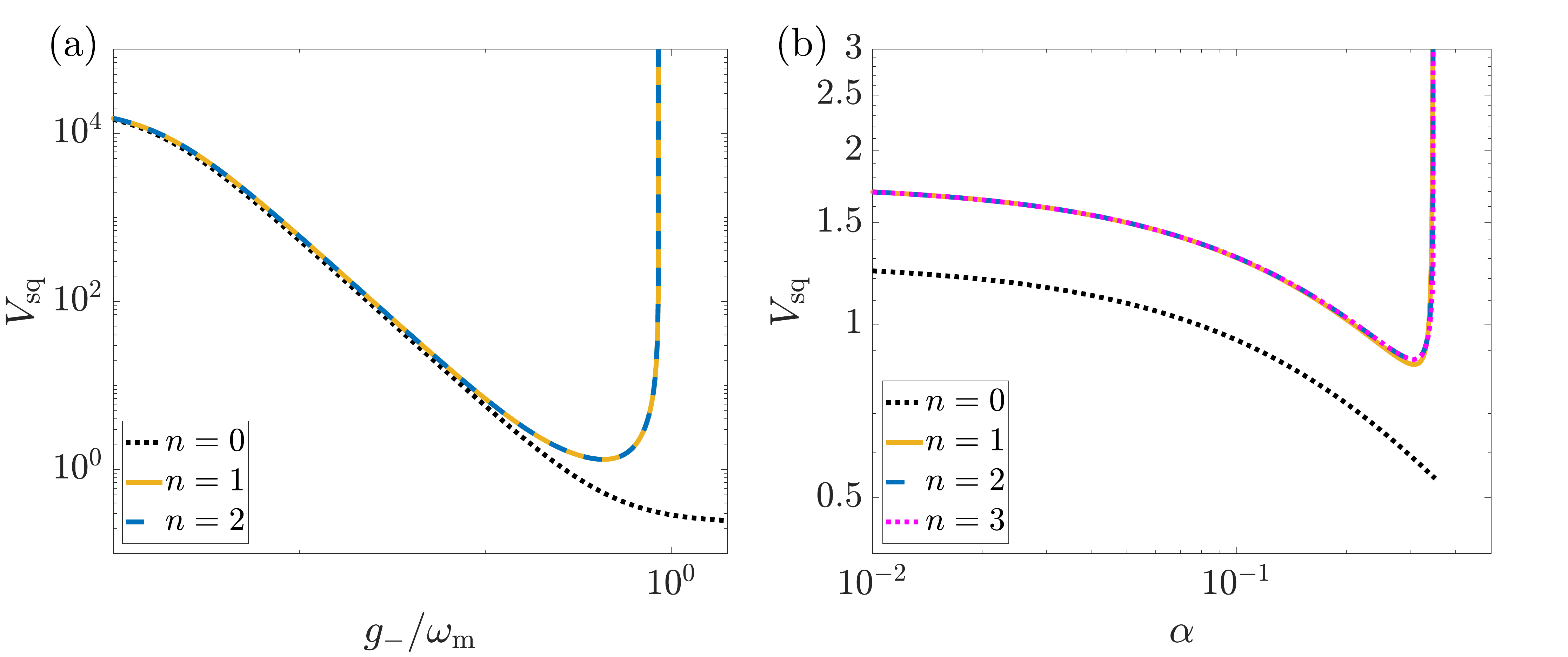}
  \caption{Squeezed variance simulated with different number of Brillouin zones for (a) the two-tone driven system of Sec.~\ref{ssec:DisSq} and (b) the levitated particle discussed in Sec.~\ref{ssec:levitation}. The parameters used in the simulations are (a) $g_+/g_-=0.7$, $\kappa/\omega_{\rm m}=0.7$, $\gamma/\omega_{\rm m}=2\times 10^{-6}$, and $\bar{n}=10^4$, (b) $g/\omega_{\rm m}=0.5$, $\kappa/\omega_{\rm m}=0.7$, $\gamma/\omega_{\rm m}=10^{-9}$, and $\bar{n}=2\times 10^7$.}\label{fig:conv}
\end{figure}

A crucial aspect of numerical simulations using the Floquet--Lyapunov technique is choosing suitable cutoff frequency in the infinite-dimensional Floquet space.
For the simulations in this manuscript, we confirm this convergence with respect to the number of Brillouin zones by repeating the calculations with different number of Brillouin zones as detailed in Fig.~\ref{fig:conv} where we plot the squeezed variances for dissipative squeezing via a two-tone drive (panel (a)) and for the levitated particle (panel (b)) with different numbers of Brillouin zones.

Considering only the first Brillouin zone (the frequency component $n=0$; the dotted black lines in Fig.~\ref{fig:conv}) corresponds to the \RWA{}.
For the two-tone driven dissipative squeezing (Fig.~\ref{fig:conv}(a)), adding the next two Brillouin zones (corresponding to the sine and cosine components for the frequency component $n=1$) already seems sufficient as this limit agrees perfectly with the next level of the approximation (\emph{i.e.}, adding the two components for the frequency $n=2$).
For the simulations in Sec.~\ref{ssec:DisSq}, we therefore use five Brillouin zones (frequency components $n\leq 2$) as shown explicitly in the drift matrix~\eqref{eq:drift2tone}.

With the joint dissipative and parametric generation for a levitated particle, more Brillouin zones have to be considered as the interaction Hamiltonian includes higher frequency components;
here, agreement is reached one step later in the approximation between the cases of $n\leq 2$ and $n\leq 3$ as shown in Fig.~\ref{fig:conv}(b).
For the simulations in Sec.~\ref{ssec:levitation}, we use the first seven Brillouin zones (frequencies $n\leq 3$), corresponding to the drift matrix~\eqref{eq:driftLev}.

%

\end{document}